\documentclass[prd,twocolumn,nopacs,floatfix,amsmath,amssymb,floatfix,nofootinbib]{revtex4}
\usepackage{subfigure,graphicx,dcolumn,booktabs,bm}
\usepackage{longtable,lscape}
\usepackage{txfonts}
\usepackage{overpic}
\usepackage{amssymb}
\usepackage{indentfirst}
\usepackage{feynmf}
\usepackage{slashed}
\usepackage{cases}
\usepackage{color,ulem}
\usepackage{multirow}
\usepackage{epstopdf}
\usepackage{makecell}
\usepackage{threeparttable}
\usepackage[
			colorlinks,
			pdfborder=001,
			citecolor=blue]{hyperref}

\begin{document}
\title{Searching for $ X_{0}(2900) $ and $ X_{1}(2900) $ through the kaon induced reactions}

\author{Qing-Yong Lin$^{1}$\footnote{corresponding author}}\email{qylin@jmu.edu.cn}
\author{Xiao-Yun Wang$^{2}$}\email{xywang@lut.edu.cn}
\affiliation{
	$^1$Department of Physics, School of Science, Jimei University, Xiamen 361021, China\\
	$^2$Department of physics, Lanzhou University of Technology, Lanzhou 730050, China
}

\date{\today}

\begin{abstract}
The productions of the exotic states $ X_0(2900) $ and $ X_1(2900) $ in  $K^{+}p \to \Sigma_{c}^{++}X_{0}(2900)$ and $K^{+}p \to \Sigma_{c}^{++}X_{1}(2900)$ reactions are studied within an effective Lagrangian approach. The $t-$channel Born term is included to calculate the production cross sections. The numerical results show that the total cross section may reach up to about 0.5 nb and 25 nb for the production of $ X_0(2900) $ and $ X_1(2900) $, respectively. The predicted differential cross sections are sensitive to the scattering angles and have strong forward enhancements. The cross sections for the $K^{+}n \to \Lambda_{c}^{+}X_{0,1}$ are also calculated and found to be larger than that in $K^{+}p$ collisions by more than one order of magnitude. The results are helpful to the possible experimental measurements with kaon beam.
\end{abstract}
\maketitle

\section{Introduction}\label{sec1}

In the past two decades, amounts of new hadronic states were discovered experimentally in the $ B $-meson decay processes. The explorations for these new particles have broaden our understanding on the nature of hadrons. However, a substantial part of these particles cannot easily find a position in the predicted conventional hadron spectrum. People speculate these states might be the missing exotic hadron states expected from QCD. In the year 2020, LHCb reported two exotic states $ X_0(2900) $ and $ X_1(2900) $ by reconstructing the $ D^-K^+ $ invariant mass in the $ B^+\to D^+D^-K^+ $ decay channel with model-depend \cite{LHCb:2020pxc} and model-independent \cite{LHCb:2020bls} assumptions. The masses and widths of $ X_{0,1}(2900) $ were determined to be
\begin{eqnarray*}
	X_0(2900):& M=2866 \pm 7 \pm 2 \text{MeV}, \\
		& \Gamma=57 \pm 12 \pm 4 \text{MeV};\\
	X_1(2900):& M=2904 \pm 5 \pm 1 \text{MeV}, \\
		& \Gamma=110 \pm 11 \pm 4 \text{MeV}.
\end{eqnarray*}
And the spin parity quantum numbers of $ X_0(2900) $ and $ X_1(2900) $ are fitted to be $ 0^{+} $ and  $ 1^{-} $, respectively. Up to now, knowledge is still sparse on $ X_{0,1}(2900) $. Appearance of the $ D^- $ and $ K^+ $ mesons in the final state implies that both of $ X_{0,1}(2900) $ can strongly couple to $ D^-K^+ $. If interpreted as resonances, they should be constituted of at least four quarks $ (ud\bar{s}\bar{c}) $ and regarded as the first observation of exotic hadrons with open flavour and without heavy quark-antiquark pairs. Once comfirmed, it should help to deepen our understanding of the QCD confinement.

Before the release of the LHCb announcement, several explorations have been carried out to discussed the existence of hadrons containing a charm and a strange quark through a coupled channel unitary approach  using the extended local hidden gauge model \cite{Molina:2010tx}, chromomagnetic interaction model \cite{Liu:2016ogz} and color-magnetic interaction model \cite{Cheng:2020nho}, in which a lowest-lying state at a mass around 2.85 GeV was predicted.

Subsequently, the discovery of $ X_{0,1}(2900) $ has attracted a great deal of attention and various identifications are proposed to explain the nature of $ X_{0,1}(2900) $. One of the mainstream views is that $ X_{0,1}(2900) $ are treated as diquark-antidiquark states or hadronic molecules. With the QCD sum rule method \cite{Wang:2020xyc,Zhang:2020oze} and a phenomenological model \cite{Karliner:2020vsi}, the authors regarded $ X_0(2900) $ as the ground-state scalar tetraquark $ sc\bar{u}\bar{d} $. While an assumption was made that $ X_0(2900) $ and $ X_1(2900) $ may be the radial excitation and orbitally excitation of $ ud\bar{s}\bar{c} $ tetraquark state, respectively \cite{He:2020jna}. The authors modeled the $ X_0(2900) $ \cite{Wang:2020prk} and $ X_1(2900) $  \cite{Agaev:2020nrc,Agaev:2021knl} as an exotic scalar and vector state built of $ ud\bar{s}\bar{c} $ and got the mass and width consistent with the LHCb experiment. However, the authors pointed that the compact $ ud\bar{s}\bar{c} $ tetraquark in $ 0^+ $ state disfavors $ X_0(2900) $ in an extended relativized quark model \cite{Lu:2020qmp}. As for the molecular picture, the $ X_0(2900) $ and $ X_1(2900) $ are assigned as $ \bar{D}^{*}K^{*} $ molecule states in overwhelming majority of investigations \cite{Hu:2020mxp,Liu:2020nil,Chen:2020aos,Huang:2020ptc,Xue:2020vtq,Molina:2020hde,Chen:2021erj,Kong:2021ohg,Wang:2021lwy,Mutuk:2020igv}. In Refs. \cite{Qi:2021iyv,He:2020btl,Chen:2021tad}, the authors investigated $ X_1(2900) $ as $ \bar{D}_1 K $ molecular state. However, there exists different opinion that does not support the $ X_1(2900) $ as molecular state \cite{Liu:2020nil,Huang:2020ptc,Dong:2020rgs}. Although the $ X_{0,1}(2900) $ can be well described with compact tetraquark and molecular pictures, other interpretations such as hadronic rescattering effects still cannot be ruled out \cite{Liu:2020orv,Burns:2020epm,Burns:2020xne}. More discussions can be found in Ref. \cite{Chen:2022asf}.

In other words, there are different and controversial interpretations of the structures of $ X_0(2900) $ and $ X_1(2900) $ in literature. That means our understanding of the nature of the $ X_{0,1}(2900) $ is sitll incomplete. More detailed investigations are required with larger data samples or different decay modes. One notices that $ X_0(2900) $ and $ X_1(2900) $ are observed in $ B $ meson decay. Searching these two states in other production channels should implement complementary measurements and help to further understand these two states. In fact, several investigations have been carried out. For example, production of the $ X_{0,1}(2900) $ in the $ \Lambda_b $ and $ \Xi_b $ decays were suggested in Ref. \cite{Hsiao:2021tyq}. The authors in Ref. \cite{Abreu:2020ony} investigated the hadronic effects on the $ X_{0,1}(2900) $ in heavy-ion collisions. Besides the two processes, one may expect the production of the $ X_{0,1}(2900) $ in the kaon induced reactions due to the strong coupling between the $ X_{0,1}(2900) $ and $ D^-K^+ $. The reaction of kaon and proton is an ideal process to explore the new hadronic states \cite{Obraztsov:2016lhp,Velghe:2016jjw,Nagae:2008zz,Wang:2019uwk,Wang:2022sib}. Thus in the present work, we explore the production of the $ X_{0,1}(2900) $ in the processes $ K^+ p \to \Sigma_c^{++} X_{0,1} $ and $ K^+ n \to \Lambda_c^{+} X_{0,1} $ with $ t- $channel $ D^- $ exchange, where the effective Lagrangian approach and Regge model are adopted. Along the way, the cross sections and angular momentum distributions are discussed.

The paper is organized as follows. After the Introduction, we explore the production mechanism of the $ X_{0,1}(2900) $ which are induced by the kaon in Sec. \ref{sec2}. In Sec. \ref{sec3}, the numerical results are given and discussed. This paper ends with a summary.

\section{Production relevant to $ X_0 $ and $ X_1 $ induced by a kaon}\label{sec2}

\begin{figure}[htbp]
\center
\includegraphics[width=0.6\linewidth]{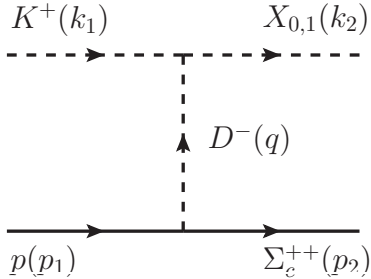}
\caption{The diagram responsible for the production of $ X_0 $ and $ X_1 $ in $K^+ p$ collisions, where the definitions of the kinematics are also shown.}\label{fig:kpto2}
\end{figure}

The tree level Feynman diagrams for the production of $ X_0 $ and $ X_1 $ are depicted in Fig. \ref{fig:kpto2}, where the $ t $-channel $ D^- $ exchange is considered. In the present work, the contributions from the $ u $ and $ s $ channels are ignored because they are strongly suppressed due to the reason that an additional $ s\bar{s} $ quark pair creation is created. Hence, the $ K^+ p \to X \Sigma_c^{++} $ reaction should be dominated by the Born term through the $ t $-channel.

\subsection{Lagrangians}

To compute the processes as shown in Fig. \ref{fig:kpto2}, the following effective Lagrangian densities are introduced as\cite{He:2011jp,Dong:2010xv,Xie:2015zga,Huang:2016ygf,Haglin:1999xs,Azevedo:2003qh}
\begin{eqnarray}
	\mathcal{L}_{DN\Lambda_c} &=& ig_{DN\Lambda_c}\bar{N} \gamma_5 D\Lambda_c + \rm{H.c.}, \\
	\mathcal{L}_{DN\Sigma_c} &=& -ig_{D^0N\Sigma_c}\bar{N} \gamma_5 D\bm{\tau}\cdot\bm{\Sigma_c} + \rm{H.c.},\\
	\mathcal{L}_{KDX_0} &=& - g_{KDX_0} \bar{X}_{0}KD + \rm{H.c.}, \\
	\mathcal{L}_{KDX_1} &=& i g_{KDX_1} \bar{X}_1^\mu(K\partial_\mu D - \partial_\mu KD) + \rm{H.c.},
\end{eqnarray}
where $N$, $ D $ and $ K $ represent the isospin doublets for necleon, the pseudoscalar $ D $ and $ K $ mesons, respectively. $ \Lambda_c $ and $ \Sigma_c $ denote the $ \Lambda_c(2286) $ and $ \Sigma_c(2455) $ isospin triplet, respectively. In the following formulae, the abbreviations $g_{\Lambda_c}\equiv g_{DN\Lambda_c}$, $g_{\Sigma_c}\equiv g_{DN\Sigma_c}$, $ g_{0}\equiv g_{KDX_0} $ and $g_{1}\equiv g_{KDX_1}$ are implemented. The coupling constants $ g_{\Lambda_c}=-13.98 $ and $ g_{\Sigma_c}=-2.69 $ are determined from the SU(4) invariant Lagrangians in terms of $ g_{\pi NN}=13.45 $ and $ g_{\rho NN}=6.0 $ \cite{Dong:2010xv,Xie:2015zga}. The coupling constants $ g_{0} $ and $ g_{1} $ will be discussed later.

It should be noted that the above effective Lagrangians depend on the quantum number rather than the internal components of $ X_{0,1} $. The information on the internal structure is coded in the coupling constants $ g_0 $ and $ g_1 $, which can be inputted from different models assigned to the $ X_{0,1} $. In addition, the experimental information can be used to determine $ g_0 $ and $ g_1 $ by comparing with our results. Thus the dimensionless coupling constants $ g_0 $ and $ g_1 $ can be related to the corresponding decay widths
\begin{eqnarray}
	\Gamma_{X_0 \to K^+D^-} &=& \frac{g_0^2 |\bm{k}_K|}{8\pi m_X^2} \\
	\Gamma_{X_1 \to K^+D^-} &=& \frac{g_1^2 |\bm{k}_K|}{6\pi m_X^2}\lambda^2(m_X,m_K,m_D),
\end{eqnarray}
where $ m_X $, $ m_K $ and $ m_D $ are the masses of $ X_{0,1} $, $ K^+ $ meson and $ D^- $ meson, respectively. $ \bm{k}_K $ is the three momentum of the $ K^+ $ meson in the $ X_{0,1} $ rest frame. $ \lambda(a,b,c)=\sqrt{a^4+b^4+c^4-2(a^2b^2+b^2c^2+c^2a^2)}/(2a) $ is the K\"{a}llen function. Unfortunately, the partial decay width of the $ X_{0,1} \to D^- K^+ $ process is still unknown from the LHCb measurements. However, it is expected that the branching fraction of the $ K^+D^- $ mode is large since it is the only two-body strong decay mode observed. Here, we make an approximation that the partial decay widths are the same as the total width of $ X_{0,1} $ resonances. Then the two coupling constants are computed to be $ g_0=4.07~\text{GeV} $ and $ g_1=6.53 $. We also notice that partial decay width $ \Gamma_{X_1 \to K^+D^-} $ was obtained through the QCD sum rules and the diquark antidiquark assignment \cite{Agaev:2021knl}. The obtained value of $ g_1 $ is twice as ours when a same decay width is adopted. These results will be utilized in the following calculations.

\subsection{Feynman amplitudes}

With the Lagrangians given above, the Born amplitudes corresponding to the diagrams in Fig. \ref{fig:kpto2} can be expressed as
\begin{eqnarray}
	\mathcal{M}_{0} &=& g_0\overline{u}(p_2) \mathcal{C}_{a} u(p_1)\mathcal{P}_{D}^{F}F^2(q^2), \label{eq:famp0} \\
	\mathcal{M}_{1} &=& g_1\overline{u}(p_2) \mathcal{C}_{a} u(p_1)(q^\mu-k_1^\mu)\mathcal{P}_{D}^{F}\varepsilon_\mu^{*}F^2(q^2) \label{eq:famp1}
\end{eqnarray}
with $\mathcal{C}_{a}=\sqrt{2}g_{\Sigma_c}\gamma_5$ and $ \mathcal{P}_{D}^{F} = 1/(q^2-m_D^2) $ the Feynman propagator of the $ D $ meson. To take into account the finite extension of the relevent hadrons, we also introduce the form factor in the amplitude with the monopole form 
\begin{equation}
	F(m_{q}^{2},q^2)=\frac{\Lambda^2-m_{q}^2}{\Lambda^2-q^2}.
\end{equation}
The cutoff in the form factor can be parametrized as $\Lambda=m_{q}+\alpha\Lambda_{QCD}$ with $\Lambda_{QCD}=220$ MeV. The parameter $\alpha$ will be discussed in further detail below.

Finally, the differential cross section in the center of mass (c.m.) frame is written as
\begin{eqnarray}
	\frac{\text{d}\sigma}{\text{d}\cos\theta} = \frac{1}{32\pi s}\frac{|\bm{k}_2|}{|\bm{k}_1|}\left(\frac{1}{2}\sum_\lambda|\mathcal{M}|^2\right),
\end{eqnarray}
with $ s=(k_1+p_1)^2 $ and $ \theta $ the angle of the outgoing $ X_0/X_1 $ meson relative to the kaon beam direction in the c.m. frame. $ \bm{k}_1 $ and $ \bm{k}_2 $ are the three-momentum of the initial kaon and the final $ X_0/X_1 $, respectively.

\section{Numerical results}\label{sec3}

\begin{figure}[htb]
	\begin{center}
		\scalebox{0.9}{\includegraphics[width=\columnwidth]{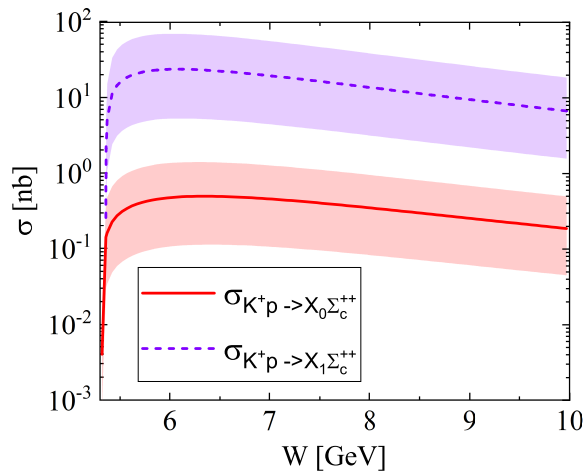}}
		\caption{The energy dependence of the total cross section for the productions of the $ X_0 $ and $ X_1 $ in $K^+ p$ collisions with $\alpha=1.5 \pm 0.5$ GeV. The red solid  and violet dashed lines are for the productions of $ X_0 $ and $ X_1 $, respectively. The bands stand for the error bar of the cutoff $\alpha$.}\label{fig:kptoxtcs}
	\end{center}
\end{figure}

With the formalism and ingredients given above, the total cross section versus the c. m. energy $W$ for $K^{+}p \to \Sigma_{c}^{++}X_{0,1}$ is calculated. The theoretical numerical results are presented in Fig. \ref{fig:kptoxtcs}. In these calculations, the parameter $\alpha$ in the parameterized cutoff is still a free parameter. It has been discussed in some literatures. For instance, by fitting the experimental data, a value of $\alpha=2.2$ was extracted for the exchanged particles $D^{*}$ and $D$ \cite{Cheng:2004ru}. However, $\alpha$ is expected to be of order unity and it depends both on the exchanged particle and the external particles involved in the strong interaction vertex. Since we do not calculate $\alpha$ from first-principles, in the present work we just employ a typical value of $\alpha=1.5$ lying between the unity and 2.2 to study the production rate.

In Fig. \ref{fig:kptoxtcs}, it is easy to find that the line shape of the cross section of $K^{+}p \to \Sigma_{c}^{++}X_{0}$ goes very rapidly near the threshold and has a maximum $0.5$ nb around $W=6.3$ GeV. It indicates that $W=6.3$ GeV is the best energy window to the search for $X_{0}$ via kaon induced reaction. The cross section for $K^{+}p \to \Sigma_{c}^{++}X_{1}$ reaction is also presented. The evolutionary trend of the line shape is similar to that of $X_{0}$ production. The total cross section reaches a maximum of about $24$ nb at $W=6$ GeV. It's obvious that the total cross section for  $K^{+}p \to \Sigma_{c}^{++}X_{1}$ is about two order of magnitude larger than that for the $X_{1}$ production.

Besides the cross sections with fixed $\alpha$, the effect of different $\alpha$ on the cross section is also studied with a range of $\alpha=1.0-2.0$. The corresponding cross sections with these different $\alpha$ values are depicted as the bands in Fig. \ref{fig:kptoxtcs}. The lower and upper edges of the band correspond to $\alpha=1.0$ and $\alpha=2.0$, respectively. That means the cross section increases with the increasing of the $\alpha$ value in the presented region.

\begin{figure}[htb]
	\begin{center}
		\scalebox{0.9}{\includegraphics[width=\columnwidth]{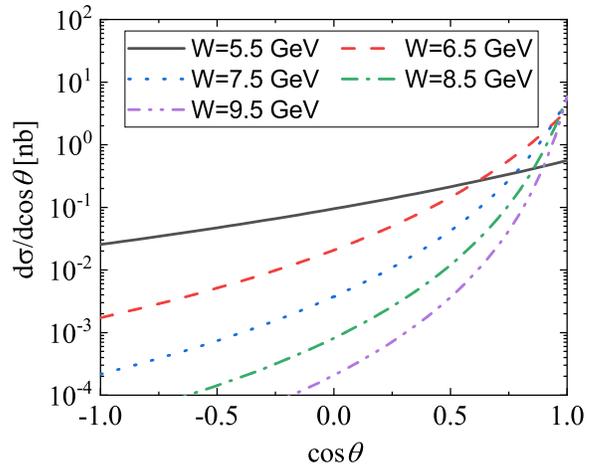}}
		\caption{The differential cross section $\text{d}\sigma/\text{d}\cos{\theta}$ of the $ X_0 $ production in $K^{+}p$ collisions at different center-of-mass energies with $\alpha=1.5$ GeV.}\label{fig:kptox0dcs}
	\end{center}
\end{figure}

\begin{figure}[htb]
	\begin{center}
		\scalebox{0.9}{\includegraphics[width=\columnwidth]{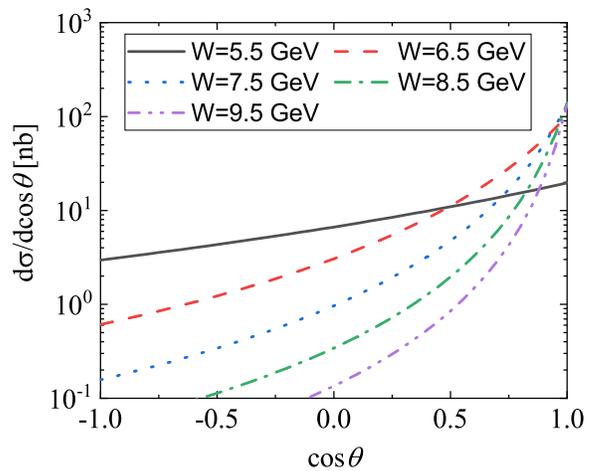}}
		\caption{The differential cross section $\text{d}\sigma/\text{d}\cos{\theta}$ of the $ X_1 $ production in $K^{+}p$ collisions at different center-of-mass energies with $\alpha=1.5$ GeV.}\label{fig:kptox1dcs}
	\end{center}
\end{figure}

In Fig. \ref{fig:kptox0dcs} and Fig. \ref{fig:kptox1dcs}, the differential cross sections of the reactions $K^{+}p \to \Sigma_{c}^{++}X_{0}$ and $K^{+}p \to \Sigma_{c}^{++}X_{1}$ are also predicted, where $\alpha=1.5$ is adopted. It is found that the differential cross sections are very sensitive to the $\theta$ angle. It's obvious that with the increase of energy, the strong forward-scattering enhancement becomes more and more apparent. Thus the experimental measurement at forward angles is highly recommended.

\begin{figure}[htb]
	\centering
	\scalebox{0.9}{\includegraphics[width=\columnwidth]{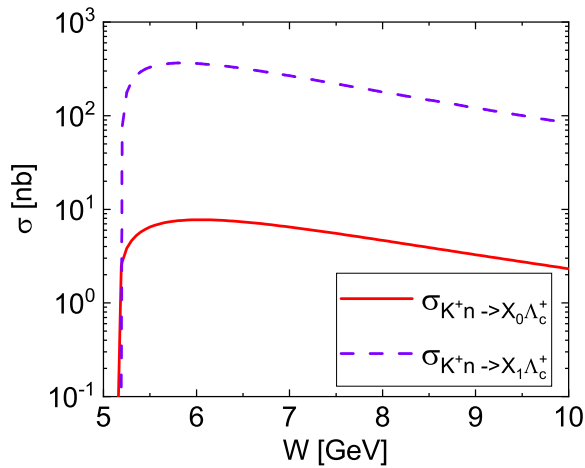}}
	\caption{The energy dependence of the total cross section for the productions of the $ X_0 $ and $ X_1 $ in $K^+ n$ collisions with $\alpha=1.5$ GeV.}\label{fig:kntoxtcs}
\end{figure}

\section{Summary}\label{sec4}

In this work, we study the productions of the exotic mesons $X_{0}(2900)$ and $X_{1}(2900)$ through the kaon induced reactions $K^{+}p \to \Sigma_{c}^{++}X_{0}$ and $K^{+}p \to \Sigma_{c}^{++}X_{1}$. With the assumption for the decay widths, the related cross sections are estimated to reach about 0.5 nb and 25 nb for the productions of $X_{0}$ and $X_{1}$, respectively. In addition, the corresponding best energy windows for searching these two states are $W=6.3$ GeV and $W=6.0$ GeV, respectively. Hence, an experimental study of $X_{0}$ and $X_{1}$ via $K^{+}p$ reaction is suggested. Moreover, a further analysis indicates that the differential cross section is quite sensitive to the $\theta$ angle and gives a considerable contribution at forward angles. These results will provide an important basis for  studying the production mechanism of $X_{0}$ and $X_{1}$ in the future. 

Besides the proton target, the neutron and deuteron targets should be alternative with kaon beam to study the productions of $X_{0}$ and $X_{1}$. In the present work, we also explore the production of $X_{0,1}$ through the process $K^{+}n \to \Lambda_{c}^{+}X_{0,1}$ with $t-$channel $D^{-}$ exchange. The relative amplitudes can be directly obtained from Eqs. \eqref{eq:famp0} and \eqref{eq:famp1} with the replacement of $\mathcal{C}_{a}$ by $\mathcal{C}_{b}=-g_{\Lambda_c}\gamma_5$. The numerical results are shown in Fig. \ref{fig:kntoxtcs} with the fixed $\alpha=1.5$. It's obvious that the cross section of $X_{0} (X_{1})$ in $K^{+}n$ reaction is about one order of magnitude larger than that in $K^{+}p$. The main reason is that the coupling of $DN\Lambda_{c}$ is much stronger than that of $DN\Sigma_{c}$. Our result is helpful to the possible experimental research of $X_{0}$ and $X_{1}$ at the kaon beam facilities such as J-PARC \cite{Aoki:2021cqa} and so on.

In addition, one notes that the spin partners of the $ X_0(2900) $, e.g., the $ J^P=1^+ $ and $ J^P=2^+ $ S-wave hadronic molecules were predicted in theories \cite{Hu:2020mxp,Liu:2020nil,Molina:2020hde,He:2020btl}. Some authors also suggest to search $ Z^{++} $, which is a doubly charged exotic state with inner structure of $ K^+D^+(cu\bar{s}\bar{d}) $ predicted in Ref. \cite{Azizi:2021aib,Agaev:2021jsz}. Thus, it is an interesting topic to study these predicted states via kaon induced reactions in the future.

\section*{Acknowledgements}
This project is supported by the Natural Science Foundation of Fujian Province under Grant No. 2018J05007 and the National Natural Science Foundation of China under Grants No. 12105115 and No. 12065014. X. Y. Wang acknowledge the West Light Foundation of
The Chinese Academy of Sciences, Grant No. 21JR7RA201.


\end{document}